\documentclass[conference]{IEEEtran}
\usepackage{cite}

\ifCLASSINFOpdf
   \usepackage[pdftex]{graphicx}
\else
\fi
\usepackage{amsmath}
\usepackage{fixltx2e}
\providecommand{\citep}{}
\renewcommand{\citep}[1] {\cite{#1}}

\providecommand{\citet}{}
\renewcommand{\citet}[1] {\cite{#1}}

\providecommand{\IEEEauthorrefmark}{}
\renewcommand{\IEEEauthorrefmark}[1] {$^{#1}$}

\usepackage{enumitem}
\usepackage{balance}
\usepackage{color}

\usepackage{kotex}
\usepackage{algorithm}
\usepackage{algpseudocode}

\newcommand{\bszero}{{\boldsymbol{0}}} 

\newcommand{\bstheta}{{\boldsymbol{\theta}}}

\DeclareSymbolFont{bbold}{U}{bbold}{m}{n}
\DeclareSymbolFontAlphabet{\mathbbold}{bbold}

\usepackage{multicol}
\usepackage{framed,color}
\definecolor{shadecolor}{rgb}{1,0.8,0.3}
\usepackage{setspace}
\begin{document}
%
\title{Near-Data Processing for\\Differentiable Machine Learning Models}



%
\author{\IEEEauthorblockN{
Hyeokjun Choe\IEEEauthorrefmark{1},
Seil Lee\IEEEauthorrefmark{1},
Hyunha Nam\IEEEauthorrefmark{1},
Seongsik Park\IEEEauthorrefmark{1},
Seijoon Kim\IEEEauthorrefmark{1},
Eui-Young Chung\IEEEauthorrefmark{2}, 
Sungroh Yoon\IEEEauthorrefmark{1,3}$^\ast$
}
\IEEEauthorblockA{
	\IEEEauthorrefmark{1}Electrical and Computer Engineering, Seoul National University, Seoul 08826, Korea
}
\IEEEauthorblockA{\IEEEauthorrefmark{2}Electrical and Electronic Engineering, 	Yonsei University, Seoul 03722, Korea
}
\IEEEauthorblockA{\IEEEauthorrefmark{3}Neurology and Neurological Sciences, Stanford University, Stanford, CA 94305, USA
}
	$^\ast$Email: sryoon@snu.ac.kr
}


\maketitle

\begin{abstract}
Near-data processing (NDP) refers to augmenting memory or storage with processing power. Despite its potential for acceleration computing and reducing power requirements, only limited progress has been made in popularizing NDP for various reasons. Recently, two major changes have occurred that have ignited renewed interest and caused a resurgence of NDP. The first is the success of machine learning (ML), which often demands a great deal of computation for training, requiring frequent transfers of big data. The second is the popularity of NAND flash-based solid-state drives (SSDs) containing multicore processors that can accommodate extra computation for data processing. In this paper, we evaluate the potential of NDP for ML using a new SSD platform that allows us to simulate in-storage processing (ISP) of ML workloads. Our platform (named ISP-ML) is a full-fledged simulator of a realistic multi-channel SSD that can execute various ML algorithms using data stored in the SSD. To conduct a thorough performance analysis and an in-depth comparison with alternative techniques, we focus on a specific algorithm: stochastic gradient descent (SGD), which is the \textit{de facto} standard for training differentiable models such as logistic regression and neural networks. We implement and compare three SGD variants (synchronous, Downpour, and elastic averaging) using ISP-ML, exploiting the multiple NAND channels to parallelize SGD. In addition, we compare the performance of ISP and that of conventional in-host processing, revealing the advantages of ISP. Based on the advantages and limitations identified through our experiments, we further discuss directions for future research on ISP for accelerating ML.

\end{abstract}


%
\IEEEpeerreviewmaketitle

\section{Introduction}
\label{Introduction}
\IEEEPARstart{T}he recent success of machine learning (ML) in various applications can be accredited to the availability of big data and powerful parallel processors that, together, have made it possible to train sophisticated models with numerous parameters. In the conventional memory hierarchy, training data is stored at a low level (e.g., hard disks) and must be moved upward all the way to the CPU registers. However, because increasingly large datasets are being used to train large-scale models such as deep networks~\citep{lecun2015deep,min2016deep}, the overhead incurred by the need to move data in the hierarchy becomes more salient and critically affects the overall computational efficiency and power consumption.

The idea behind near-data processing (NDP)~\citep{balasubramonian2014near} is to equip the memory or storage with intelligence (i.e., processors) and let it process the data stored therein firsthand. A successful NDP implementation would reduce the number of data transfers and the power consumption required, as well as offload some of the computational burden of CPUs. The various approaches to realizing NDP have included processing in memory (PIM)~\citep{gokhale1995processing,yitbarek2016exploring, azarkhish2016design, xuscaling, azarkhish2017neurostream, chi2016prime} and in-storage processing (ISP)~\citep{acharya1998active, kim2016storage, lee2016activesort, choi2015energy}.
Despite the potential of NDP, it has not yet been used in commercial systems. For the PIM approach, there has been a wide performance gap between the separate processes to manufacture logic and memory chips. For the ISP approach, commercial hard disk drives (HDDs), the traditional mainstream storage devices, typically possess limited processing capabilities due to high downward price pressures.

Recently, we have seen a renewed interest in and resurgence of NDP triggered by two major factors: one on the application side and the other on the technology side: On the application side, computing- and data-intensive analytics methods are rapidly being deployed for various machine learning tasks. For instance, training deep neural networks typically requires large volumes of data to ensure their performance. Although GPUs and multicore CPUs often provide an effective means for massive computation, training data must still be stored in storage because of its size\footnote{For instance, the popular ImageNet dataset \cite{ILSVRC15} contains over 1.2 million images that require more than 200 GB of storage space.} and then transferred to the CPU/GPU level for computation. On the technology side, NAND flash-based solid-state drives (SSDs) are becoming increasingly popular and are gradually replacing HDDs in various computing sectors. To interface SSDs with the host to seamlessly replace HDDs, SSDs require some ability to run software (e.g., for address translation and garbage collection~\citep{kim2002space,gupta2009dftl}). Therefore, SSDs are often equipped with multicore processors that provide far more processing capabilities than is available in HDDs. Usually, these SSD-based processors experience considerable idle time that can be exploited for purposes other than SSD housekeeping~\citep{kim2010architecture,kim2016effective}.


Given this context, we propose a new SSD platform that allows us to simulate ISP of machine learning workloads and evaluate the potential of NDP for machine learning in ISP. Our platform, named ISP-ML, is a full-fledged system-level simulator of a realistic multi-channel SSD that can execute various machine learning algorithms using the data stored on the SSD.
To conduct a thorough performance analysis and an in-depth comparison with alternatives, we focus on describing our implementation of a specific algorithm in this paper: the stochastic gradient descent (SGD) algorithm, which is the \emph{de facto} standard for training differentiable models (such as logistic regression and deep neural networks).
Specifically, we implement three types of parallel SGD: synchronous SGD~\citep{zinkevich2010parallelized}, Downpour SGD~\citep{dean2012large}, and elastic averaging SGD (EASGD)~\citep{zhang2015deep}. We compare the performance of these parallel SGD implementations using a 10-times amplified version of MNIST~\citep{lecun1998mnist}. Furthermore, to evaluate the effectiveness of ISP-based optimization by SGD, we compare the performance of ISP-based optimization and conventional in-host processing (IHP)-based optimization.

To the best of the authors' knowledge, this work is one of the first attempts to apply NDP to a multi-channel SSD for accelerating SGD-based optimization for training differentiable models.
Our specific contributions include the following:
\begin{itemize} 
	\item We created a full-fledged ISP-supporting SSD platform called ISP-ML, which required a multi-year team effort. ISP-ML is versatile and can simulate not only storage-related functionalities of a multi-channel SSD but also NDP-related functionalities in a realistic manner. ISP-ML can execute various machine learning algorithms using the data stored in the SSD while supporting the simulation of multi-channel NAND flash SSDs to exploit data-level parallelism.
	\item We thoroughly tested the effectiveness of our platform by implementing and comparing multiple versions of parallel SGD, which is widely used for training various machine learning algorithms. We also devised a methodology that can carefully and fairly compare the performance of IHP-based and ISP-based optimization.

	
	\item We identified intriguing future research opportunities to exploit the parallelism provided by the multiple NAND channels inside SSDs. As in high-performance computing, multiple ``nodes'' (i.e., NAND channel controllers) exist for sharing workloads, but---unlike in conventional parallel computing---the communication cost is negligible (due to the negligible latency of on-chip communication). Using our platform, we envision new designs of parallel optimization and training algorithms that can exploit this characteristic, producing enhanced results.

\end{itemize}

Of particular note is that the goal of building ISP devices is not to compete with multicore CPUs or GPUs but to complement them in systems intended for processing machine-learning workloads. By improving the performance of secondary storage devices in such a system and by reducing the data transfers involved in the training of machine learning models, ISP devices are expected to improve the overall performances of systems equipped with CPUs/GPUs.

The remainder of this paper is organized as follows: In Section~\ref{Related work}, we briefly review some fundamentals of machine learning and SSDs along with related work on NDP. In Section~\ref{Methodology}, we describe the details of our proposed ISP-ML platform. Section~\ref{Experimental Results} presents our experimental results. Finally, in Section~\ref{Discussion}, we discuss the experimental results and propose future research directions.

\section{Background}
\label{Related work}

\subsection{Machine Learning as an Optimization Problem}
Machine learning is a branch of artificial intelligence that aims to provide computers with the ability to learn without being programmed explicitly~\cite{samuel1959some}. Depending on the type of feedback available for training, we can broadly classify machine learning tasks into three categories\cite{russell1995modern}: supervised learning, unsupervised learning, and reinforcement learning.

To facilitate further explanation, we briefly review the basic formulation of supervised learning, focusing only on the materials directly relevant to the present work. More in-depth reviews of machine learning can be found in \cite{bishop2006pattern,murphy2012machine,Goodfellow-et-al-2016}.

For various types of machine learning algorithms, the core concept can often be explained using the following equations~\cite{bishop2006pattern}:
\begin{align}
F(D,\bstheta) &= L(D,\bstheta) + r(\bstheta)\label{eq1}\\
\Delta\bstheta(D) &= -\eta\nabla F(D,\bstheta)\label{eq3}\\
\bstheta_{t+1} &= \bstheta_t + \Delta\bstheta(D)\label{eq2}
\end{align}
where $D$ and $\bstheta$ denote the input data and model parameters (also known as weights), respectively, and a loss function $L(D, \bstheta)$ reflects the difference between the optimal and current hypotheses. A regularization term to mitigate the over-fitting problem is denoted by $r(\bstheta)$, and the objective function $F(D, \bstheta)$ is the sum of the loss and regularization terms. The main purpose of supervised machine learning can then be formulated as finding the optimal $\bstheta$ that minimizes $F(D,\bstheta)$.

The method of (batch) gradient descent~\cite{murphy2012machine} is a first-order iterative optimization algorithm to find the minimum value of $F(D, \bstheta)$ by updating $\bstheta$ in every iteration $t$ in the direction of the negative gradient of $F(D, \bstheta)$, where $\eta$ is the learning rate. In each iteration of the gradient descent optimization, the value of the parameter $\bstheta$ is updated as follows:
\begin{align}
\bstheta_{t+1} &= \bstheta_{t} - \eta \nabla F(D,\bstheta_t)\\
&= \bstheta_{t} - \eta  \sum_i \nabla F(D_i,\bstheta_t)\label{eq5}
\end{align}
where $t$ and $i$ are the iteration index and the data sample index, respectively. Updating $\bstheta$ by gradient descent thus requires examining all the data samples (Eq.~\ref{eq5}), which is time-consuming for large data.

To reduce this computational burden, SGD computes the gradient of the parameters and updates them using only a single training sample per iteration, which often makes SGD 	suffer from undesirable convergence behavior. Between these two is the minibatch (stochastic) gradient descent method~\cite{Goodfellow-et-al-2016}, which uses multiple (but far less than the whole, e.g., 100) samples per iteration. As will be explained shortly, we employ a type of minibatch SGD in our framework, setting the size of a minibatch to the number of training samples in a NAND flash page. We name this type of minibatch SGD \textit{page-minibatch} (see Section~\ref{Methodology}).

Among these variants of gradient descent, the minibatch SGD is often the method of choice recently, and many researchers use the term SGD to refer to the minibatch SGD. Therefore, for simplicity, we also term the minibatch SGD as simply SGD in this paper unless otherwise stated.


\subsection{Parallel and Distributed SGD}\label{ss:parallelSGD}

SGD is widely employed in machine learning for its simplicity and effectiveness~\cite{Goodfellow-et-al-2016}. However, the execution time of SGD increases rapidly as the data size grows; consequently, various approaches for accelerating SGD-based training by parallelization and/or distributed computation have been proposed.

Zinkevich et al.~\citet{zinkevich2010parallelized} proposed an algorithm that implements parallel SGD in a distributed computing architecture.
However, this algorithm often suffers from excessive latency caused by the need to synchronize all the slave nodes. To overcome this weakness, Recht et al.~\citet{recht2011hogwild} proposed the lock-free Hogwild! algorithm that can update parameters asynchronously.
Hogwild! is normally implemented on a single machine with a multicore processor. Dean et al.~\citet{dean2012large} proposed the Downpour SGD for distributed computing systems by extending the Hogwild! algorithm.
While successful, Downpour SGD often fails to overcome communication bottlenecks and exhibits inefficient bandwidth usage caused by substantial data movements between computing nodes.

Recently, a new parallel SGD algorithm called EASGD was proposed~\citep{zhang2015deep}, and many EASGD-based approaches have reported its effectiveness in distributed environments. EASGD works in a master-slave setting. The basic idea behind EASGD is to let each slave maintain its own local parameter and to link the local parameters maintained in slaves with the central parameters managed by the master using the concept of an elastic force. In EASGD, the communication to update the model weights does not occur at every iteration but at every $\tau$ iterations. The update equations for the model parameters ($\bstheta_\mathrm{slave}$ for local parameters and $\bstheta_\mathrm{master}$ for central parameters) are as follows:
\begin{align}
\bstheta_\mathrm{slave} &= \bstheta_\mathrm{slave} - \alpha(\bstheta_\mathrm{slave}-\bstheta_\mathrm{master}) \label{eq6}\\
\bstheta_\mathrm{master} &= \bstheta_\mathrm{master} + \alpha(\bstheta_\mathrm{slave}-\bstheta_\mathrm{master}) \label{eq7}
\end{align}
where the user-specified parameter $\alpha$ is called the moving rate and is related to the degree to which the local parameters are allowed to fluctuate from the centralized variable, similarly to the modulus of elasticity in kinetics.

\subsection{Fundamentals of Solid-State Drives (SSDs)}

SSDs have emerged as a type of next-generation storage device that uses NAND flash memory~\citep{kim2010architecture}. SSDs have several advantages over HDDs (such as energy consumption~\cite{yoo2011ssd}, IO performance~\cite{seo2014io}, and mechanical characteristics). As shown in the right image in Figure~\ref{architecture}(a), a typical SSD consists of an SSD controller, a DRAM buffer, and a NAND flash array. The SSD controller is typically composed of an embedded processor, a cache controller, channel controllers, and host-interface logic.

The embedded processor used inside an SSD controller usually consists of 2--4 cores with a clock frequency of 300--550MHz. 
The DRAM component, controlled by the cache controller, plays the role of a cache buffer when the NAND flash array is read or written. The size of the DRAM component ranges from 512MB to 1GB; however, recently announced products (e.g., Samsung 960 Pro) are equipped with 2GB or more of DRAM.
The NAND flash array contains multiple NAND chips that can be accessed simultaneously by employing multi-channel configurations and per-channel controllers.
The host interface logic enables a connection between the host and the SSD. SATA3 has been widely used. Recently available SSDs often support PCIe and non-volatile memory express (NVMe) for high-speed host interfaces.

For seamless, drop-in replacements of HDDs by SSDs in the host, SSDs require software support due to the distinct characteristics of the NAND flash memory inside. The unit of NAND flash read and written is a page (i.e., 4--32KB), while the unit of erases is a block (64--256 pages). In addition, the reliability and durability of NAND flash cells are limited~\cite{kim2011deduplication}. To improve the performance and durability of the NAND flash array, SSDs are thus managed by software called the flash translation layer (FTL)~\cite{chung2009survey}, which performs wear-leveling and garbage collection~\citep{kim2002space,gupta2009dftl}.




\subsection{Previous Work on Near-Data Processing}

In many large-scale data analytics systems, it is critical to minimize data movement not only to avoid overall performance degradation but also to enhance power-efficiency and reliability. The paradigm of data processing in various domains is swiftly moving from computing-centric to data-centric. Inspired by these trends, the concept of NDP~\citep{balasubramonian2014near} has recently attracted considerable interest. In NDP, computation is performed in the most appropriate location (other than the CPU/GPU). For example, computation might be performed in memory or in the storage device where the input data reside~\cite{chi2016prime}. We can divide existing NDP approaches into two main categories, namely PIM and ISP.

PIM aims at performing computation inside main memory. Subsequent to the pioneering work by Gokhale et al.~\cite{gokhale1995processing}, various PIM approaches have been proposed. Recently, Yitbarek et al.~\cite{yitbarek2016exploring} reported accelerator logic for data-intensive operations (e.g., sorting, string matching, memcopy, and hash-table lookups) in a type of three-dimensional memory called a hybrid memory cube (HMC)~\cite{pawlowski2011hybrid,hmc2013hybrid}. Azarkhish et al.~\cite{azarkhish2016design} proposed another HMC architecture called smart memory cube (SMC) and verified its performance using a simulator.

The popularity of deep learning~\cite{lecun2015deep} and its heavy computation demands has sparked the development of new PIM approaches. Xu et al.~\cite{xuscaling} implemented a convolutional neural network (CNN)~\cite{Goodfellow-et-al-2016} on an HMC-based PIM. Azarkhish et al.~\cite{azarkhish2017neurostream} proposed a flexible SMC-based PIM called NeuroCluster and implemented a CNN on it. The authors of both approaches reported that they outperformed GPU-based deep learning implementations in terms of performance and power-efficiency. Chi et al.~\cite{chi2016prime} proposed a resistive random access memory (ReRAM)-based PIM architecture that employed a crossbar array to perform matrix multiplications efficiently.

Early ISP approaches include the Active Disks architecture proposed by Acharya et al.~\cite{acharya1998active}. Many existing ISP methods have focused on popular (but inherently simple) algorithms to perform scan, join, and query operations~\citep{kim2016storage}. Lee et al.~\cite{lee2016activesort} proposed running a merge operation (frequently used by external sort operations in Hadoop) inside an SSD to reduce IO transfers and read/write operations, which also functions to extend the lifetime of the NAND flash inside the SSD. Choi et al.~\cite{choi2015energy} implemented algorithms for linear regression, $k$-means, and string matching in the flash memory controller (FMC) via reconfigurable stream processors. In addition, they implemented a MapReduce application inside the embedded processor and FMC of the SSD using partitioning and pipelining methods that both improved performance and reduced power consumption. BlueDBM~\citep{jun2015bluedbm} is an ISP system architecture for distributed computing systems with a flash memory-based embedded field programmable gate array (FPGA). The authors implemented nearest-neighbor search, graph traversal, and string search algorithms. Biscuit~\cite{gu2016biscuit} is an SSD framework equipped with FMCs with pattern matching logic that enables MySQL queries.

\subsection{Our Approach}

Of the two NDP categories (PIM and ISP), the approach described here belongs to the ISP category. Compared with the existing ISP approaches, our method is unique in that it supports gradient-based training for differentiable models. To the best of the authors' knowledge, no prior work has ever implemented and evaluated ISP-based optimization of machine learning algorithms using a gradient descent algorithm such as SGD. Furthermore, our methodology supports parallel SGD, which is key for enabling large-scale training on massive data.

Besides its widespread use, SGD has some appealing characteristics that facilitate hardware implementations. We can implement parallel SGD on top of the master-slave architecture of the cache controller and the channel controllers. We can also take advantage of effective techniques initially developed for distributed and parallel computations. Importantly, each SGD iteration is not overly complex and can be implemented without incurring excessive hardware overhead.

The next section describes more details of our proposed ISP-ML platform and the implementation of parallel SGD using this platform.

\section{Proposed Methodology}
\label{Methodology}

\begin{figure*}
	\centering
	\includegraphics[width=0.95\linewidth]{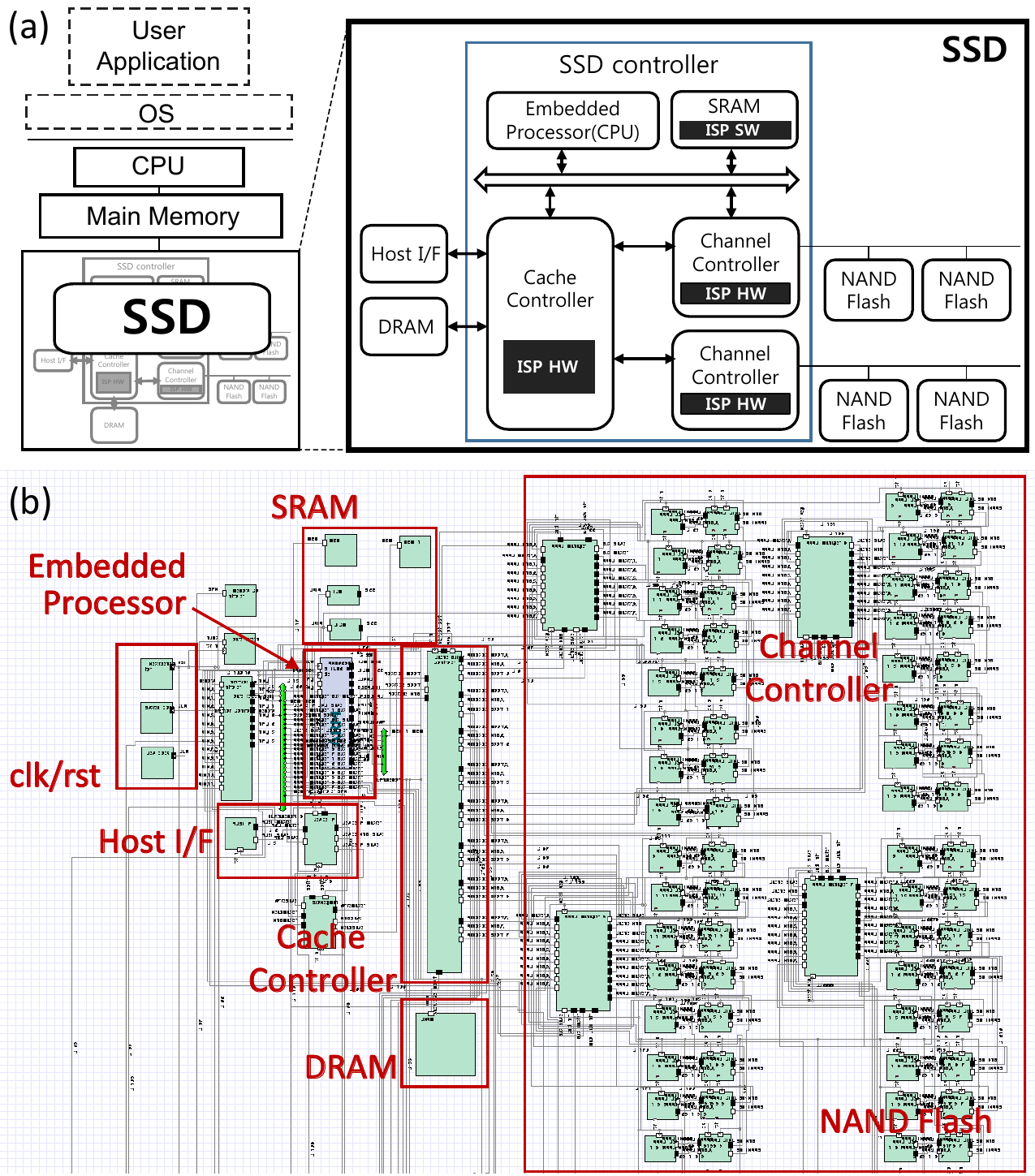}
	
	%
	\caption{(a) Block diagram of a typical computing system equipped with an SSD and a magnified view of a typical SSD depicting its internal components and their connections. (b) Schematic of the proposed ISP-ML framework, which is implemented in SystemC using Synopsys Platform Architect (http://www.synopsys.com). Refer to the main text for more details.}
	\label{architecture}
\end{figure*}

Figure~\ref{architecture}(a) shows a block diagram of a typical computing system assumed to have an SSD as its storage device. The figure also includes a magnified view of the SSD block diagram showing the major components of an SSD and their interconnections. Starting from the baseline SSD depicted above, we can implement ISP functionalities by modifying the components marked with black boxes (i.e., the ISP HW and ISP SW in the figure). Figure~\ref{architecture}(b) shows the detailed schematic of our proposed ISP-ML platform, which corresponds to the SSD block (with ISP components) shown in Figure~\ref{architecture}(a).

In this section, we provide more details concerning our ISP-ML framework. In addition, we propose a performance comparison methodology that can compare the performance of ISP and of conventional IHP in a fair manner. As a specific example of the ML algorithms that can be implemented in ISP-ML, we utilize parallel SGD.



%

\subsection{ISP-ML: ISP Platform for Machine Learning on SSDs}
Our ISP-ML is a system-level simulator implemented in SystemC on the Synopsys Platform Architect environment (http://www.synopsys.com). ISP-ML can simulate both the hardware and software ISP components marked in Figure~\ref{architecture}(b) simultaneously. This integrative functionality is crucial to design space exploration in SSD development. Moreover, ISP-ML allows us to execute various machine learning algorithms described in high-level languages (C or C++) directly on ISP-ML with only minor modifications. To yield a reasonable simulation speed, we modeled ISP-ML at a cycle-accurate transaction level while minimizing the negative impact on accuracy. Specific parameters and considerations used in our implementation can be found in Section~\ref{ss:exp-setup}.

At the conception of this research, we could not find any publicly available SSD simulator that could be modified to  implement ISP functionalities. This motivated us to implement a new simulator. There are multiple ways to implement the idea of ISP in an SSD. The first option would be to use the embedded core inside the SSD controller (Figure~\ref{architecture}(a)). This option does not require designing new hardware logic, and it is flexible because the ISP capability is implemented in software. However, this option is not ideal for exploiting hardware acceleration and parallelization. The second option would be to design dedicated hardware logic chips (such as those boxes with black marks in Figure~\ref{architecture}(a)) and integrate them into the SSD controller. Although significantly more effort is required for this latter option compared to the first option, we chose the second option because of the long-term advantages provided by hardware acceleration and power reduction.

Specifically, we implemented two types of ISP hardware components, in addition to the software components. First, we let each channel controller not only manage read/write operations to/from its NAND flash channel (as in the usual SSDs), but also perform primitive operations on the data stored in its NAND channel. The type of primitive operation performed depends on the machine learning algorithm used (the next subsection explains such operations for SGD in more detail). Additionally, each channel controller in ISP-ML (slave) communicates with the cache controller (master) using a master-slave architecture.
Second, we designed the cache controller so it can collect the outcomes from each of the channel controllers, in addition to its inherent functionality as a cache (DRAM) manager inside the SSD controller. This master-slave architecture can be interpreted as a tiny-scale version of the master-slave architecture commonly used in distributed systems. Just like the channel controllers, the exact functionality of the cache controller can be optimized depending on the specific algorithm used. Both the channel controllers and the cache controller have internal memory. Here, the memory size in the cache controller is larger to aggregate the partial results from the channel controllers.

\subsection{Parallel SGD Implementation on ISP-ML}
\setlength{\fboxsep}{0.9pt}
\begin{figure*}
\begin{multicols}{3}
\begin{algorithm}[H]
\scriptsize
\caption{ISP-Based Synchro. SGD}
\label{syncsgd}
\begin{algorithmic}[1]
\State Read page-sized data $D^i_{j}$ from NAND array
\Statex $\triangleright~i$: channel controller index
\Statex $\triangleright~j$: NAND flash page index
\Statex $\triangleright~k$: training sample index (within a minibatch)
\begin{spacing}{1.42}

\State Pull $\bstheta_\mathrm{cache}$ from the cache controller buffer
\State $\bstheta^i \gets \bstheta_\mathrm{cache}$
\State $\Delta\bstheta^i \gets \bszero, \quad k\gets 0$
\While {$k < b$} \Comment{$b$: minibatch size}
\State Calculate $F(D^i_{jk},\bstheta^i)$
\State $\Delta\bstheta^i \gets \Delta\bstheta^i + \eta \nabla F(D^i_{jk},\bstheta^i)$
\State $k\gets k+1$
\EndWhile
\State Push $\Delta\bstheta^i$ and wait
\Statex $\triangleright~$ Lines 13--14: executed by the cache controller
\State \fbox{$\bstheta_\mathrm{cache} \gets \bstheta_\mathrm{cache} - \frac{1}{n} \sum_i \Delta \bstheta^i$}
\State \fbox{Signal each channel controller}
\end{spacing}
\end{algorithmic}
\end{algorithm}

\begin{algorithm}[H]
\scriptsize
\caption{ISP-Based Downpour SGD}
\label{dpsgd}
\begin{algorithmic}[1]
\State Read page-sized data $D^i_{j}$ from NAND array
\Statex $\triangleright~i$: channel controller index
\Statex $\triangleright~j$: NAND flash page index
\Statex $\triangleright~k$: training sample index (within a minibatch)
\begin{spacing}{1.445}

\State Pull $\bstheta_\mathrm{cache}$ from the cache controller buffer
\State $\bstheta^i \gets \bstheta_\mathrm{cache}$
\State $\Delta\bstheta^i \gets \bszero, \quad k\gets 0$
\While {$k < b$} \Comment{$b$: minibatch size}
\State Calculate $F(D^i_{jk},\bstheta^i)$
\State $\Delta\bstheta^i \gets \Delta\bstheta^i + \eta \nabla F(D^i_{jk},\bstheta^i)$
\State $k\gets k+1$
\EndWhile
\If {$j ~\mathrm{mod}~ \tau = 0$} \Comment{$\mathrm{mod}$: modulus}
\State Push $\Delta\bstheta^i$
\State \fbox{$\bstheta_\mathrm{cache} \gets \bstheta_\mathrm{cache} - \Delta \bstheta^i$}\Comment{by cache ctrl.}
\EndIf
\end{spacing}
\end{algorithmic}
\end{algorithm}

\begin{algorithm}[H]
\scriptsize
\caption{ISP-Based EASGD}
\label{easgd}
\begin{spacing}{1.2}
\begin{algorithmic}[1]
\State Read page-sized data $D^i_{j}$ from NAND array
\Statex $\triangleright~i$: channel controller index
\Statex $\triangleright~j$: NAND flash page index
\Statex $\triangleright~k$: training sample index (within a minibatch)
\State $k \gets 0$
\While {$k < b$} \Comment{$b$: minibatch size}
\State Calculate $F(D^i_{jk},\bstheta^i)$
\State $\textsf{temp} \gets \textsf{temp} + \eta \nabla F(D^i_{jk},\bstheta^i)$
\State $k\gets k+1$
\EndWhile
\State $\bstheta^i \gets \bstheta ^i - \frac{1}{b} \textsf{temp}$
\If {$j ~\mathrm{mod}~ \tau = 0$} \Comment{$\mathrm{mod}$: modulus}
\State Pull $\bstheta_\mathrm{cache}$ from the cache controller buffer
\State $\textsf{temp} \gets \bstheta_\mathrm{cache}$
\State $\Delta\bstheta^i \gets \alpha(\bstheta^i-\textsf{temp})$
\State $\bstheta^i \gets \bstheta^i-\Delta\bstheta^i$
\State Push $\Delta\bstheta^i$
\State \fbox{$\bstheta_\mathrm{cache} \gets \bstheta_\mathrm{cache} + \Delta \bstheta^i$}\Comment{by cache ctrl.}
\EndIf
\end{algorithmic}
\end{spacing}
\end{algorithm}
\end{multicols}
\caption{Pseudo-code of the three SGD algorithms implemented in ISP-ML: synchronous SGD~\citep{zinkevich2010parallelized}, Downpour SGD~\citep{dean2012large}, and EASGD~\citep{zhang2015deep}. The \fbox{boxed lines} indicate the computation occurring in the cache controller (master); the other lines are executed in the channel controllers (slaves). The user-specified parameters are $b$ (minibatch size), $\eta$ (learning rate), $\tau$ (communication frequency), and $\alpha$ (moving rate). We use the so-called \textit{page-minibatch}, where the size of a minibatch is set to the number of training samples in a NAND flash page. Each channel controller (slave) \textit{pulls} the weight information from the cache controller (master) and \textit{pushes} the updated parameters to the cache controller.}
	\label{pseudo}
\end{figure*}

Using our ISP-ML platform, we implemented the three types of parallel SGD algorithms (synchronous SGD~\citep{zinkevich2010parallelized}, Downpour SGD~\citep{dean2012large}, and EASGD~\citep{zhang2015deep}). To adapt the original algorithms to our NAND-based SSD platform, we customized each of these algorithms as shown in Figure~\ref{pseudo}.
In this section, we focus on describing the details of our ISP-based implementation and the customization of these SGD algorithms and omit the purely algorithmic details of each method; we refer interested readers to the corresponding references.

Our ISP-ML platform runs the parallel SGD in a master-slave manner, where the master is the cache controller and the slaves are the channel controllers. Algorithms 1--3 (shown in Figure~\ref{pseudo}) describe the operations executed by each channel controller (the boxed operations in each algorithm are processed by the cache controller). Adopting the common terminology used in the distributed optimization literature, we assume that each channel controller (slave) \textit{pulls} the weight information from the cache controller (master) and \textit{pushes} the updated parameters to the cache controller.

Note that the size of a minibatch for the minibatch SGD in our framework is set to the number of training samples in a NAND flash page (this type of minibatch is referred to as \textit{page-minibatch} in this paper). Also note that $\bstheta_\mathrm{cache}$ refers to the (global) model parameters managed by the cache controller, while $\bstheta^i$ indicates the (local) model parameters updated by the $i$-th channel controller.


\subsubsection{\textbf{ISP-Based Synchronous SGD}}
Algorithm~\ref{syncsgd} describes our ISP-based implementation of synchronous SGD. In this scheme, each of the channel controllers computes the gradient in parallel, and the cache controller iteratively aggregates the model parameters in a synchronous manner. The pseudocode shows how the $i$-th channel controller (lines 1--10) and the cache controller (lines 11--12) operate together to fulfill the $j$-th ISP operation.

In line 1, the $i$-th channel controller reads the data $D^i_j$ from the $j$-th page of the $i$-th NAND flash channel into its buffer. This data contains $b$ training samples indexed by $k$ ($k=0,1,\ldots,b-1$), where $b$ is the size of each page-minibatch (i.e., as explained above, the number of training samples in a NAND flash page). The channel controller also pulls the cache controller's parameters ($\bstheta_\mathrm{cache}$) to set the initial value of the local parameters ($\bstheta^i$) in lines 2--3. Using the data and parameters stored in the buffer, each channel controller calculates the gradient in parallel (lines 5--9, where $D^i_{jk}$ indicates the $k$-th training sample in $D^i_j$). After sending the updated gradient to the cache controller, each channel controller waits for a signal from the cache controller (line 10). The cache controller aggregates and updates the parameters using the gradient information received from all the channel controllers (line 11). After updating $\bstheta_\mathrm{cache}$ is  complete, the cache controller signals the channel controllers to start the next iteration (line 12).




\subsubsection{\textbf{ISP-Based Downpour SGD}}

As shown in Algorithm~\ref{dpsgd}, the initial operations for Downpour SGD are similar to those in synchronous SGD (lines 1--9 are identical). 
The main difference comes after each channel controller calculates the gradient $\Delta\bstheta^i$ (lines 10--13 of Algorithm~\ref{dpsgd}).
In Downpour SGD, each channel controller communicates with the cache controller not after every minibatch update (as in the synchronous SGD) but only after every $\tau$ minibatches (line 10). That is, the user-specified $\tau$ parameter controls the frequency of master-slave communications. After sending the gradient update information ($\Delta\bstheta^i$) to the cache controller after every $\tau$-th minibatch, the channel controller does not wait for the signal from the cache controller for synchronization (as occurs in the synchronous SGD implementation) but immediately starts the next minibatch. The cache controller updates the $\bstheta_\mathrm{cache}$ parameters sequentially using the gradients from the channel controllers.




\subsubsection{\textbf{ISP-Based EASGD}}
As reviewed in Section~\ref{ss:parallelSGD}, in EASGD, each slave maintains its own parameters (unlike synchronous SGD and Downpour SGD in which slaves compute gradient information to update the central parameters but do not maintain local parameters). Therefore, as shown in Algorithm~\ref{easgd}, the channel controller does not start by pulling the central parameters $\bstheta_\mathrm{cache}$. Instead, each channel controller updates its own $\bstheta^i$ parameters in lines 2--8. Similar to Downpour SGD, after processing every $\tau$-th minibatch, each channel controller adjusts its parameters by communicating with the cache controller (lines 9--16).
Each channel controller pulls the parameters from the cache controller (line 10), calculates the differences between its own parameters and the cache controller's parameters (line 13), and then pushes the differences to the cache controller (line 14). Note that lines 13 and 15 in Algorithm~\ref{easgd} correspond to Eq.~\ref{eq6} and Eq.~\ref{eq7} as discussed in Section~\ref{ss:parallelSGD}, respectively. The cache controller updates its parameters $\bstheta_\mathrm{cache}$ using the information received from the channel controllers in line 15.

\subsection{Methodology for IHP-ISP Performance Comparison}\label{ss:method-ihp-comparison}

\begin{figure*}
	\centering
	\includegraphics[width=0.965\linewidth]{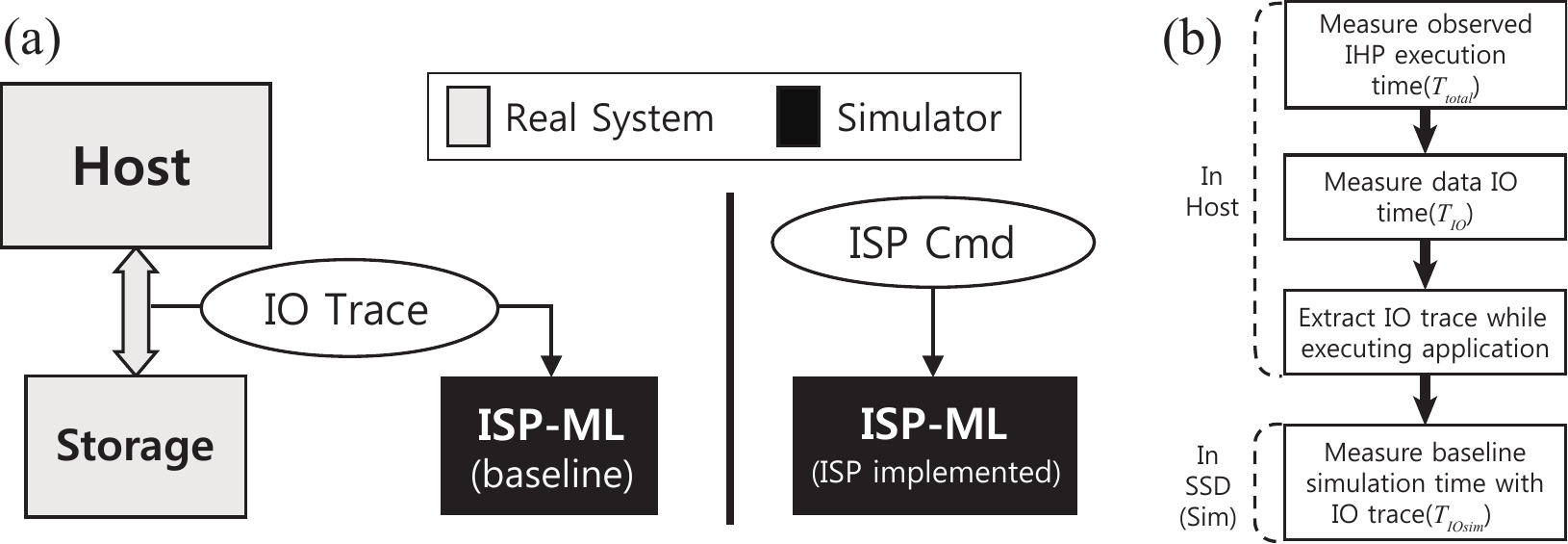}
	\caption{(a) Overview of our methodology to compare the performance of in-host processing (IHP) and in-storage processing (ISP). (b) Details of our IHP-ISP comparison flow.}
	\label{comparisonMethod}
\end{figure*}

To evaluate the effectiveness of ISP, it is crucial to accurately and fairly compare the performances of ISP and conventional IHP. However, performing this type of comparison is not trivial (see Sections~\ref{ss:result-ihp-comparison} and ~\ref{ss:discussion-b} for additional discussion of this topic). Furthermore, accurately modeling commercial SSDs equipped with ISP-ML is impossible due to a lack of information about the commercial SSDs (e.g., there is no public information on the FTL or internal architecture of any commercial SSD). Therefore, we propose a practical methodology to accurately compare IHP and ISP performances, which is depicted in Figure~\ref{comparisonMethod}. Note that this comparison methodology is applicable not only to the parallel SGD implementations explained above but also to other ML algorithms that could be executed in ISP-ML.


In the proposed comparison methodology, we focus on the data IO latency time of the storage (denoted as $T_\mathrm{IO}$), because the latency is the most critical factor among those that affect the execution time of IHP. The observed IHP execution time ($T_\mathrm{total}$) can then be divided into data IO time and non-data IO time ($T_\mathrm{nonIO}$) as follows:
\begin{align}
\textrm{Observed~IHP~execution~time} = T_\mathrm{total} = T_\mathrm{nonIO} + T_\mathrm{IO}.
\label{equation:ihpTime}
\end{align}

To calculate the expected IHP simulation time adjusted to ISP-ML, the data IO time of IHP is replaced by the data IO time of the baseline SSD in ISP-ML ($T_\mathrm{IOsim}$). Using Eq.~(\ref{equation:ihpTime}), the expected IHP simulation time can then be represented by
\begin{align}
\textrm{Expected~IHP~simulation~time} &= T_\mathrm{nonIO} + T_\mathrm{IOsim}\\ &= T_\mathrm{total} - T_\mathrm{IO} + T_\mathrm{IOsim}.
\label{equation:eIhpTime}
\end{align}

The overall flow of the proposed comparison methodology is depicted in Figure~\ref{comparisonMethod}(b). First, the total processing time ($T_\mathrm{total}$) and the data IO time of storage ($T_{IO}$) are measured in IHP, extracting the IO trace of storage during an application execution. The simulation IO time ($T_\mathrm{IOsim}$) is then measured using the IO trace (extracted from IHP) on the baseline SSD of ISP-ML. Finally, the expected IHP simulation time is calculated by plugging the total processing time ($T_\mathrm{total}$), the data IO time of storage ($T_\mathrm{IO}$) and the simulation IO time ($T_\mathrm{IOsim}$) into Eq.~(\ref{equation:eIhpTime}). With the proposed method and ISP-ML, which is applicable to a variety of IHP environments regardless of the type of storage used, it is possible to quickly and easily compare the performances of various ISP implementations and IHP in a simulation environment.

\section{Experimental Results}
\label{Experimental Results}
\subsection{Setup and Implementation}\label{ss:exp-setup}
All the experiments presented in this section were executed on a computer equipped with an 8-core Intel(R) Core i7-3770K CPU (3.50GHz) with 32GB of DDR3 RAM running Ubuntu 14.04 LTS (kernel version: 3.19.0-26-generic) and a Samsung SSD 840 Pro. We used an ARM 926EJ-S (400MHz) as the embedded processor inside ISP-ML and DFTL~\citep{gupta2009dftl} as the FTL for ISP-ML.

The NAND flash simulation model and its parameters used in our experiments were derived from a commercial product (Micron NAND MT29F8G08ABACA) and had the following specifications: page size $=$ 8KB, $t_\mathrm{prog} =$ 300$\mu$s, $t_\mathrm{read} = $75$\mu$s, and $t_\mathrm{block~erase} = $ 5$m$s.\footnote{These are conservative settings compared with those of the original commercial product; using the specifications of a commercial product will thus improve the performance of ISP-ML.}

Each channel controller had 24KB of memory [8KB (page size) for data and 16KB for ISP] and a floating point unit (FPU) with a 0.5 instruction/cycle performance (with pipelining). The cache controller was equipped with memory of $(n+1)\times $ 8KB (page size), where $n$ is the number of channels ($n=4,8,16$). Depending on the algorithm running in ISP-ML, we can adjust these parameters.

%


The main purpose of the experiments described in this paper was to verify the functionality of our ISP-ML framework and to evaluate the effectiveness of ISP over conventional IHP using SGD, even though our framework is certainly not limited to SGD. To this end, we selected logistic regression, a fundamental ML algorithm that can directly show the advantage of ISP-based optimizations over IHP-based optimizations without unnecessary complications. We implemented the logistic regression algorithm as a single-layer perceptron (with cross-entropy loss) in SystemC and uploaded it to ISP-ML. As stated in Section~\ref{ss:future-work}, our future work includes the implementation and testing of more complicated models (such as deep neural networks) to capitalize on the opportunities for improvement revealed from the experiments presented in this paper.

As the nonlinear activation function~\cite{Goodfellow-et-al-2016} of the perceptron model, we used the sigmoid function $s(t)=1/(1+\exp(-t))$. Because the FPU we used did not support an exponential operation, we designed custom logic to implement the $\exp(\cdot)$ function based on previous work~\cite{namin2009efficient}. Our implementation could compute the sigmoid function in one cycle (2.5ns) with a maximum error of 0.04.


As test data, we utilized the samples from the MNIST database~\citep{lecun1998mnist}. To amplify the number of training samples and show the scalability of our approach, we used elastic distortion~\citep{simard2003best} to produce 10 times more data than the original MNIST (approximately 600,000 training and 10,000 test samples were used in total). To focus on the performance evaluation of running ISP operations, we preloaded our NAND flash simulation model with the simulation data (the same condition was used for the alternatives for fairness). Based on the size of a training sample in this dataset and the size of a NAND page (8KB), we set the size of each minibatch to 10.





\subsection{Performance Comparison: ISP-Based Optimization}
\begin{figure}
	\centering
	\includegraphics[width=0.95\linewidth]{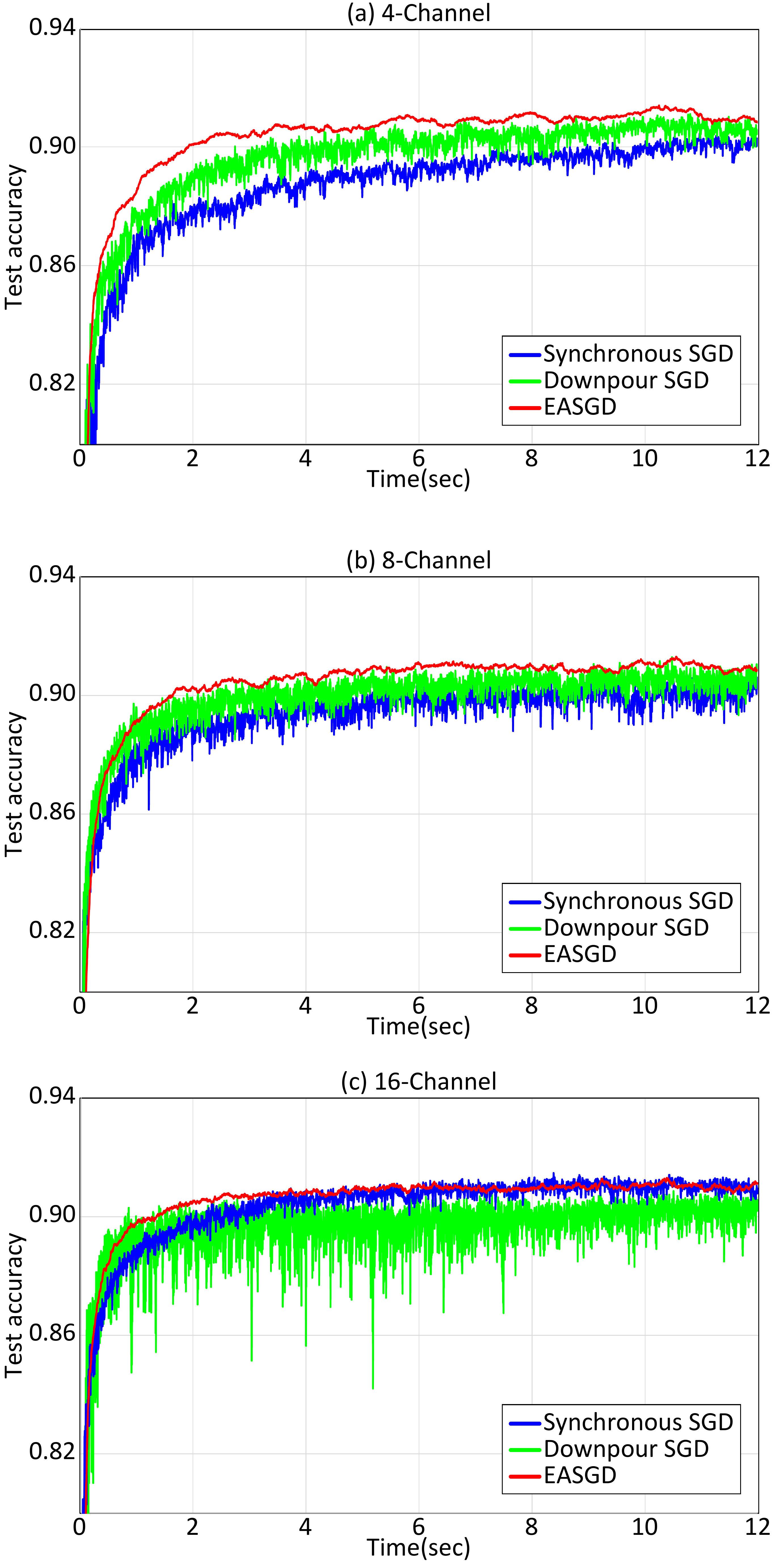}
	\caption{Test accuracy of three ISP-based SGD algorithms versus wall-clock time with a varying number of NAND flash channels: (a) 4 channels, (b) 8 channels, and (c) 16 channels.}
	\label{comparisonISP}
\end{figure}


As previously explained, to identify which SGD algorithm would be best suited for use in ISP, we implemented and analyzed three types of SGD algorithms: synchronous SGD, Downpour SGD, and EASGD. For Downpour SGD and EASGD, we set the communication period ($\tau$) to 1. The moving rate ($\alpha$) for EASGD was 0.001. To perform a fair comparison, we chose different learning rates for different algorithms that yielded the best performance for each algorithm. Figure~\ref{comparisonISP} shows the test accuracy of three algorithms with varying numbers of channels (4, 8, and 16) with respect to wall-clock time.




As shown in Figure~\ref{comparisonISP}, EASGD obtained the best convergence speed in all the cases tested. On average, EASGD outperformed synchronous and Downpour SGD by factors of 2.96 and 1.41, respectively.
%
In the case of 4 and 8 channels, synchronous SGD showed a slower convergence speed when compared to Downpour SGD because it could not begin learning on the next set of mini-batch until all the channel controllers had reported their results to the cache controller. For 16 channels, synchronous SGD showed faster convergence speed than Downpour SGD. A reasonable explanation for this result would be that Downpour SGD calculates gradients based on more outdated parameters as the number of channels increases~\citep{chen2016revisiting}.
This result suggests that EASGD is adequate for all the channel configurations tested in the sense that ISP can benefit from ultra-fast on-chip level communication and employ application-specific hardware that can eliminate interruptions from other processors.





\subsection{Performance comparison: IHP versus ISP}\label{ss:result-ihp-comparison}
\begin{figure*}
	\centering
	\includegraphics[width=0.85\linewidth]{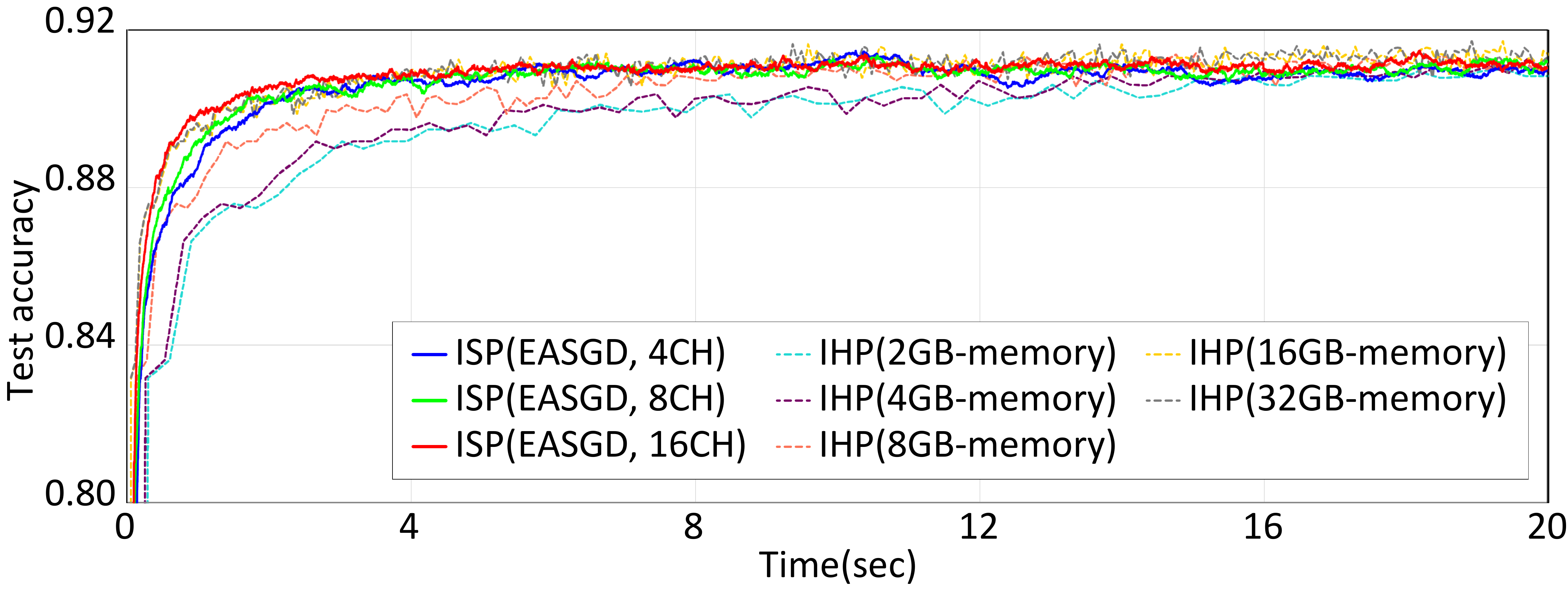}
	\caption{Test accuracy of ISP-based EASGD in the 4, 8, and 16 channel configurations and IHP-based minibatch SGD using diverse memory sizes.}
	\label{ISPvsIHP}
\end{figure*}




In large-scale machine learning, the computing systems used may suffer from memory shortages, which incur significant data swapping overhead. In this regard, ISP can provide an effective solution that can potentially reduce the penalty from data transfer by processing core operations at the storage level.

In this context, we carried out additional experiments to compare the performance of IHP-based and ISP-based EASGD. We tested the effectiveness of ISP in a memory shortage situation with 5 different configurations of IHP memory: 2GB, 4GB, 8GB, 16GB, and 32GB. We assumed that the host had already loaded all the data to main memory for IHP. This assumption is realistic because state-of-the-art machine learning techniques often employ a prefetch strategy to hide the initial data transfer latency.


As depicted in Figure~\ref{ISPvsIHP}, ISP-based EASGD with 16 channels obtained the best performance in our experiments. The convergence speed of the IHP-based optimization slowed down in accordance with the reduced memory size. The results with 16GB and 32GB of memory yielded similar results because 16GB was sufficient to load and allocate most of the resources required by the process. Consequently, ISP was more efficient when memory was insufficient, as would often be the case with large-scale datasets in practice.


\begin{figure*}
	\centering
	\includegraphics[width=0.95\linewidth]{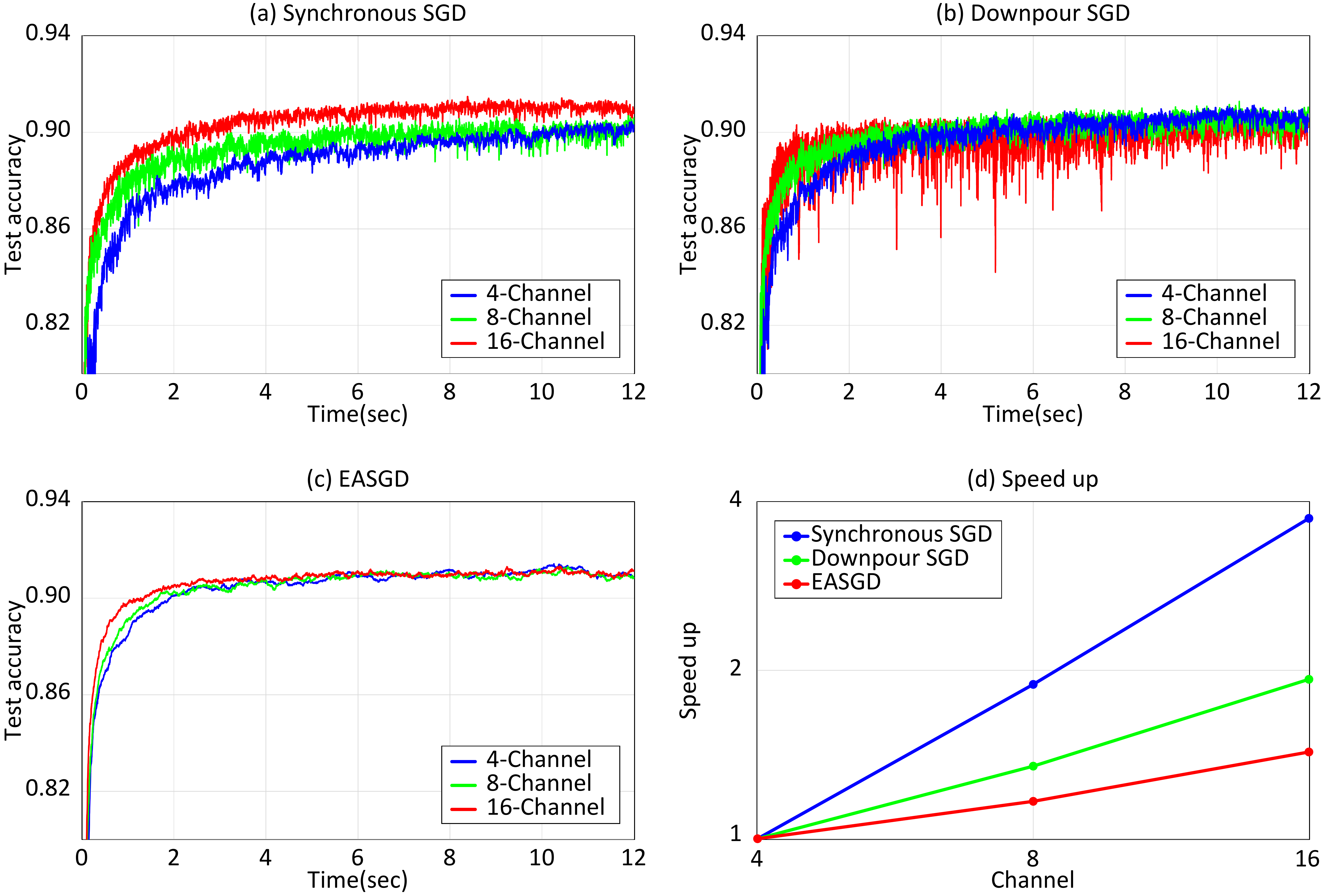}
	\caption{Test accuracy of different ISP-based SGD algorithms for a varied number of channels: (a) synchronous SGD, (b) Downpour SGD, and (c) EASGD. (d) Training speed improvements for the three SGD algorithms for varying numbers of channels.}
	\label{comparisonChannel}
\end{figure*}


\subsection{Channel Parallelism}
To closely examine the effect of exploiting data-level parallelism on performance, we compared the accuracy of the three SGD algorithms when varying the number of channels (4, 8, and 16), as shown in Figure~\ref{comparisonChannel}. The convergence speed of all three algorithms improved by using more channels; synchronous SGD improved by 1.93 times when the number of channels increased from 8 to 16. As shown in Figure~\ref{comparisonChannel}(d), the improvement in convergence speed tends to be proportional to the number of channels. These results suggest that the communication overhead in ISP is negligible and that ISP does not suffer from the  communication bottleneck that commonly occurs in distributed computing systems.

\begin{figure*}
	\centering
	\includegraphics[width=0.923\textwidth]{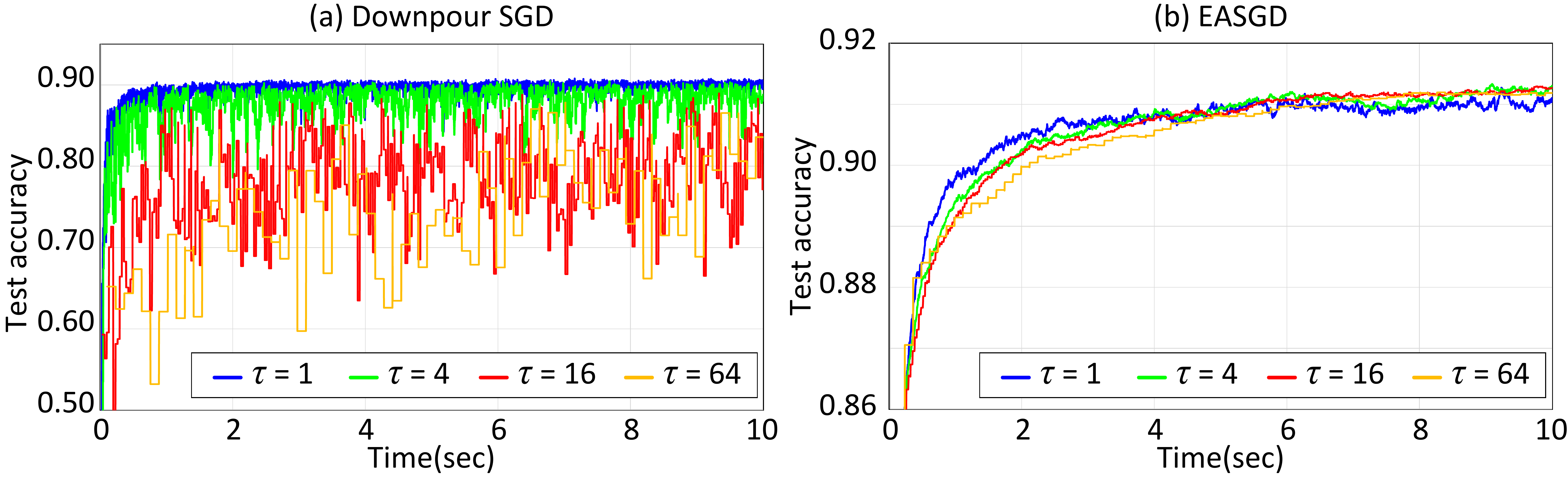}
	\caption{Test accuracy of ISP-based Downpour SGD and EASGD algorithms versus wall-clock time for different communication periods.}
	\label{comparisonCommPeri}
\end{figure*}


\subsection{Effects of Communication Period in Asynchronous SGD }

Finally, we investigated how changes in the communication period (i.e., how often data exchange occurs during distributed optimization) affect SGD performance in the ISP environment. Figure~\ref{comparisonCommPeri} shows the test accuracy of the Downpour SGD and EASGD algorithms versus wall-clock time when we varied their communication periods. As described in \citet{zhang2015deep}, Downpour SGD normally achieved a high performance for a low communication period [$\tau = 1,4$] and became unstable for a high communication period [$\tau = 16,64$] in ISP. Interestingly, in contrast to the conventional distributed computing system setting, the performance of EASGD decreased as the communication period increased in the ISP setting. This result occurs because the on-chip communication overhead in ISP is significantly lower than that in the distributed computing system. As a result, there is no need to extend the communication period to reduce the communication overhead in the ISP environment.


\section{Discussion}
\label{Discussion}
\subsection{Parallelism in ISP}
Given the advances in underlying hardware and semiconductor technology, ISP can provide various advantages for the types of data processing involved in machine learning. For example, our ISP-ML platform could minimize (practically eliminate) the overhead of communication between parallel nodes leveraged by ultra-fast on-chip communication inside an SSD. Minimizing communication overhead can improve various key aspects of data-processing systems, such as energy efficiency, data management, security, and reliability. 
By exploiting the advantages of fast on-chip communications in ISP, we envision that we will be able to devise a new kind of parallel algorithm for optimization and machine learning running on ISP-based SSDs.

The results of our experiments also revealed that a high degree of parallelism could be achieved by increasing the number of channels inside an SSD. Some of the currently available commercial SSDs have as many as 16 channels. Given that the commercial ISP-supporting SSDs would (at least initially) be targeted at high-end SSD markets with many NAND flash channels, our approach is expected to add valuable functionality to such SSDs. Unless carefully optimized, a conventional distributed system will see diminishing returns as the number of nodes increases, due to the increased communication overhead and other factors. Exploiting a hierarchy of parallelism (i.e., parallel computing nodes, each of which has ISP-based SSDs with parallelism inside) may provide an effective acceleration scheme, although a fair amount of additional research is needed before we can realize this idea.

\subsection{ISP-IHP Comparison Methodology}\label{ss:discussion-b}
To fairly compare the performances of ISP and IHP, it would be ideal to implement ISP-ML in a real semiconductor chip, or to simulate IHP in the ISP-ML framework. However, selecting either option is possible but not plausible in an academic setting because of high chip manufacturing costs, and the prohibitively high simulation time to simulate IHP in the Synopsys Platform Architect environment (we would have to implement many components of a modern computer system to effectively simulate IHP). Another option would be to implement both ISP and IHP using FPGAs, but doing so will require another round of significant development efforts.

To overcome these challenges  while still assuring a fair comparison between ISP and IHP, we proposed the comparison methodology described in Section~\ref{ss:method-ihp-comparison}. In terms of measuring the absolute running time, our methodology is not ideal. However, for revealing the relative performance between alternatives, our method provides a satisfactory solution.

Our comparison methodology extracts IO trace from the storage while executing an application in the host. This trace is then used to measure the simulation IO time in the baseline SSD in ISP-ML. In this procedure, we assume that the non-IO time of IHP is consistent regardless of the type of storage the host has. The validity of this assumption is warranted by the fact that the amount of non-IO time changed by the storage is usually negligible compared with the total execution time or IO time. 





\subsection{Opportunities for Future Research}\label{ss:future-work}


In this paper we focused on implementing and testing ISP-based SGD as a proof of concept. The simplicity and popularity of (parallel) SGD underlie our choice. By design, it is possible to run other algorithms in our ISP-ML framework immediately; recall that our framework includes a general-purpose ARM processor that can run executables compiled from C/C++ code. However, it would be meaningless to have an ISP-based implementation if its performance were unsatisfactory. To unleash the full power of ISP, we need additional ISP-specific optimization efforts, as is typically the case with hardware design.
%
%

With this in mind, we have started implementing deep neural networks (with  realistic numbers of layers and hyperparameters) using our ISP-ML framework. In particular, we are carefully devising a way of balancing the memory usage in the DRAM buffer, the cache controller, and the channel controllers inside ISP-ML. It is reasonable to envision an SSD with a DRAM cache consisting of a few gigabytes of memory; in contrast, it is unrealistic to design a channel controller with that much memory. Given that a large amount of memory is needed only to store the parameters of such deep models, and that IHP and ISP have different advantages and disadvantages, it would be intriguing to investigate how to make IHP and ISP cooperate to enhance the overall performance. For instance, we could let ISP-based SSDs perform low-level data-dependent tasks while assigning high-level tasks to the host, expanding the current roles of the cache controller and the channel controllers inside ISP-ML to the whole system level.

Our future work also includes the following: First, we will be able to implement adaptive optimization algorithms such as Adagrad~\citep{duchi2011adaptive} and Adadelta~\citep{zeiler2012adadelta}. Second, pre-computing metadata during data writes (instead of data reads) is another research direction that could result in even greater performance improvements. Third, we will be able to implement data-shuffling functionality to maximize the effect of data-level parallelism. Currently, ISP-ML arbitrarily splits the input data into its multi-channel NAND flash array. Fourth, we may investigate the effect of NAND flash design on performance, such as the NAND flash page size. Typically, the size of a NAND flash page significantly affects the performance of SSDs, given that the page size (e.g., 8KB) is the basic unit of NAND operation (read and write). In cases where the size of a single example exceeds the page size, frequent data fragmentation is inevitable, which will eventually affect the overall performance. The effectiveness of using multiple page sizes has already been reported for conventional SSDs~\citep{kim2016nand}. We may borrow this idea to further optimize ISP-ML.

\section*{Acknowledgment}
The authors would like to thank Byunghan Lee and other members of Data Science Laboratory, Seoul National University, and Design Technology Laboratory, Yonsei University for constructive discussions. This work was supported by ICT R\&D program of MSIP/IITP. (No.R7117-16-0235), by the National Research Foundation of Korea (NRF) funded by the Ministry of  Science, ICT \& Future Planning (2014M3C9A3063541, 2016M3A7B4911115), and by the Brain Korea 21 Plus Project in 2017.



%
%



\bibliographystyle{IEEEtran}
%
%
%

\balance
\bibliography{iclr2017_conference,msst2017}

\end{document}